\documentclass[letterpaper]{article} % DO NOT CHANGE THIS
\usepackage{amsfonts}
\usepackage{aaai2026}  % DO NOT CHANGE THIS
\usepackage{times}  % DO NOT CHANGE THIS
\usepackage{helvet}  % DO NOT CHANGE THIS
\usepackage{courier}  % DO NOT CHANGE THIS
\usepackage[hyphens]{url}  % DO NOT CHANGE THIS
\usepackage{graphicx} % DO NOT CHANGE THIS
\usepackage{natbib}  % DO NOT CHANGE THIS AND DO NOT ADD ANY OPTIONS TO IT
\usepackage{caption} % DO NOT CHANGE THIS AND DO NOT ADD ANY OPTIONS TO IT
\usepackage{algorithm}

\usepackage{cite}
\usepackage{amsmath,amssymb,amsfonts}
\usepackage{graphicx}
\usepackage{textcomp}
\usepackage{bm}
\usepackage{subfigure}
\usepackage{url,multirow}

\usepackage{cite}
\usepackage{amsmath,amssymb,amsfonts}
\usepackage{graphicx}
\usepackage{textcomp}
\usepackage{amsmath}
\usepackage{cases}
\usepackage{diagbox}
\usepackage{booktabs}
\usepackage{graphicx}
\usepackage{subfigure}
\usepackage{algpseudocode}
\usepackage{amsmath}
\usepackage{graphics}
\usepackage{epstopdf}
\usepackage{bm}
\usepackage{amssymb,amsfonts}
\usepackage{color}
\usepackage{float}
\usepackage{lineno}

\usepackage{newfloat}
\usepackage{listings}
\DeclareCaptionStyle{ruled}{labelfont=normalfont,labelsep=colon,strut=off} % DO NOT CHANGE THIS
\floatstyle{ruled}
\newfloat{listing}{tb}{lst}{}
\floatname{listing}{Listing}
\title{REFS: Robust EEG feature selection with missing multi-dimensional annotation for emotion recognition}
\author{
    \textbf{Xueyuan Xu*}, \textbf{Wenjia Dong}, \textbf{Fulin Wei}, \textbf{Li Zhuo} \\
    {\normalfont School of Information Science and Technology, Beijing University of Technology, Beijing 100124, China} \\
    {\normalfont School of Artificial Intelligence, Anhui University, Beijing 100124, China} \\
    {\normalfont \{xxy, zhuoli\}@bjut.edu.cn, 23027425@emails.bjut.edu.cn, weifulin@ahu.edu.cn}
}

\usepackage{bibentry}
\begin{document}

\maketitle

\begin{abstract}
The affective brain-computer interface is a crucial technology for affective interaction and emotional intelligence, emerging as a significant area of research in the human-computer interaction. Compared to single-type features, multi-type EEG features provide a multi-level representation for analyzing multi-dimensional emotions. However, the high dimensionality of multi-type EEG features, combined with the relatively small number of high-quality EEG samples, poses challenges such as classifier overfitting and suboptimal real-time performance in multi-dimensional emotion recognition. Moreover, practical applications of affective brain-computer interface frequently encounters partial absence of multi-dimensional emotional labels due to the open nature of the acquisition environment, and ambiguity and variability in individual emotion perception. To address these challenges, this study proposes a novel EEG feature selection method for missing multi-dimensional emotion recognition. The method leverages adaptive orthogonal non-negative matrix factorization to reconstruct the multi-dimensional emotional label space through second-order and higher-order correlations, which could reduce the negative impact of missing values and outliers on label reconstruction. Simultaneously, it employs least squares regression with graph-based manifold learning regularization and global feature redundancy minimization regularization to enable EEG feature subset selection despite missing information, ultimately achieving robust EEG-based multi-dimensional emotion recognition. Simulation experiments on three widely used multi-dimensional emotional datasets, DREAMER, DEAP and HDED, reveal that the proposed method outperforms thirteen advanced feature selection methods in terms of robustness for EEG emotional feature selection.
\end{abstract}


\section{Introduction}
Based on the strategy of depicting emotions, emotion representation models can be broadly classified into two categories: discrete emotional models and multi-dimensional emotional models. Compared to discrete emotional models, multi-dimensional emotional models offer broader characterization and are capable of describing the evolution of emotional states \cite{ezzameli2023emotion}. Electroencephalography (EEG) is a non-invasive technology that monitors neural activities and can quickly respond to diverse emotional states \cite{yang2024improving,zhang2024beyond}. Recently, EEG-based affective brain-computer interface (a-BCI) systems have garnered significant attention in the fields of affective interaction and emotional intelligence due to its high temporal resolution and undisguisable characteristics \cite{si2024eeg,liu2024enhancing}. Multiple EEG feature extraction approaches, including differential entropy, function connectivity, and absolute power etc, have been employed to analyze the non-stationary and nonlinear specificities of EEG recordings, thereby facilitating precise emotional state identification.

\begin{figure}[!t]
\centering
\includegraphics[width=0.456\textwidth]{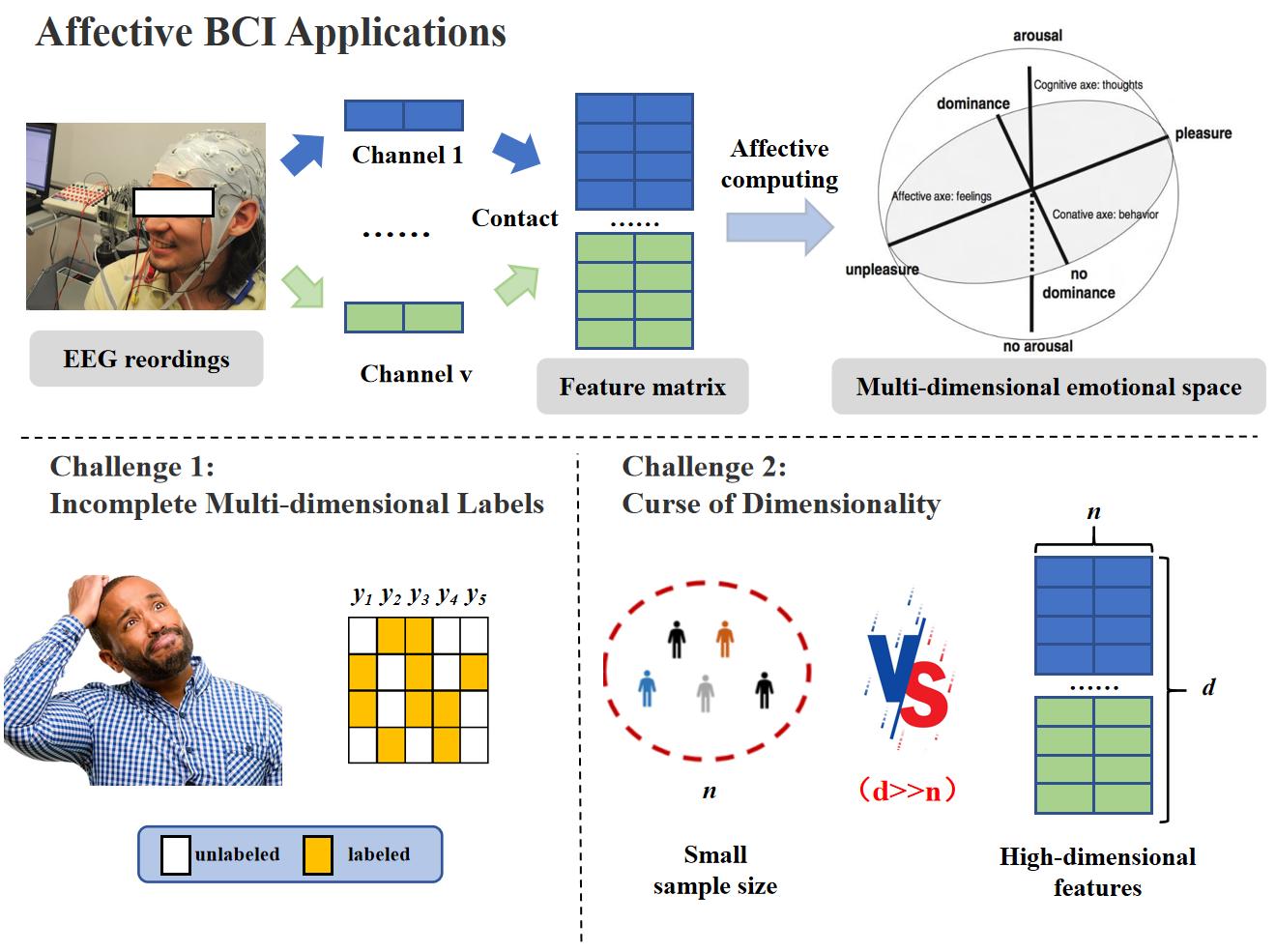}
\caption{An illustration of the affective BCI applications, along with two challenges: (1) Incomplete multi-dimensional emotional labels induced by the open nature of the acquisition environment, and ambiguity and variability in individual emotion perception; (2) The high dimensionality of multi-type EEG features and the relatively small number of samples lead to the curse problem of dimensionality.}
\label{Illustration}
\end{figure}

The number of sensors available for affective computing is growing rapidly owing to advancements in EEG acquisition devices, leading to a vast amount of features derived from these electrodes \cite{wang2020taffc}. However, given the relatively small number of high-quality EEG instances, the features often become high-dimensional and may include redundant, useless, or interference information, which can substantially impair identification performance of emotional states \cite{wang2020emotion,GRMOR2021taffc}. Feature selection is a valid approach for identifying significant features and eliminating irrelevant ones from the original set, thereby preserving the essential neural representation of EEG features and improving the transparency and interpretability of the affective computing model \cite{xu2024wsel,jenke2014taffc}. According to their feature evaluation and search mechanisms, EEG feature selection methods can be broadly categorized into following three types: filter, wrapper, and embedded methods \cite{zhang2019review}. 

Filter methods evaluate the significance of EEG features in emotion recognition based on the statistical or information entropy characteristics of the data. However, these techniques often yield unsatisfactory EEG feature selection performance, regardless of the interactions the learning models \cite{zhang2019review}. To solve this problem, several studies have explored wrapper methods \cite{nakisa2018evolutionary,asemi2025improving}, which adopt random or sequential search strategies to "wrap" features into candidate subsets and then a learning or prediction model is employed to compare the performance of these candidate subsets. The wrapper techniques generally obtain better learning performance than filter methods. Nonetheless, wrapper methods often involve numerous trials and incur substantial computational costs \cite{saeys2007review}. 

Recently, there has been growing interest in embedded approaches as a potential alternative to overcome the limitations of filter methods. Embedded techniques integrate the search for informative and non-redundant EEG features into the model optimization problem, evaluating the relative importance of each EEG feature while optimizing the learning models. Their effectiveness in EEG-based affective computing tasks has been demonstrated \cite{xu2020fsorer, GRMOR2021taffc}.

Existing EEG feature selection studies usually assume that EEG emotional data are complete. Nevertheless, due to the open nature of the acquisition environment, and ambiguity and variability in individual emotion perception, practical applications of the a-BCI frequently encounter partial absence of multi-dimensional emotional labels. This missing information directly hinders accurate modeling of the relationship between multi-channel EEG recordings and multi-dimensional emotional labels.

To address the challenges, we employ extended adaptive orthogonal non-negative matrix factorization to propose a robust EEG feature selection (REFS) method with incomplete multi-dimensional emotion recognition, as illustrated in Fig.\ref{Illustration}. This method reconstructs the label space by leveraging second-order and higher-order correlations within the multi-dimensional emotional labels, while simultaneously selecting discriminative and non-redundant EEG features despite the presence of incomplete multi-dimensional emotional labeling information.

Furthermore, the contributions of our work are as follows:

\begin{itemize}
  \item[$\bullet$] An incomplete multi-dimensional emotional feature selection method is proposed for a-BCI applications. REFS integrates EEG emotional feature selection into a extended adaptive orthogonal non-negative matrix factorization framework. It can effectively mitigates the impact of missing values and outliers on EEG feature selection model construction, and incorporates second-order and higher-order correlations within multi-dimensional emotional labels to restore the missing emotional labels.
  \item[$\bullet$] To address the optimization problem of REFS, an efficient alternative scheme is proposed to guarantee convergence and achieve an optimal solution.
  \item[$\bullet$] To validate the effectiveness of REFS for incomplete multi-dimensional emotion feature selection, we employed three public EEG-based emotion datasets, DREAMER, DEAP, and HDED, which contain multi-dimensional emotional labels. Experimental results show that REFS outperforms thirteen state-of-the-art feature selection methods, achieving superior emotion recognition performance across six performance metrics.
\end{itemize}

\begin{figure*}[!t]
\centering
\includegraphics[width=0.9\textwidth]{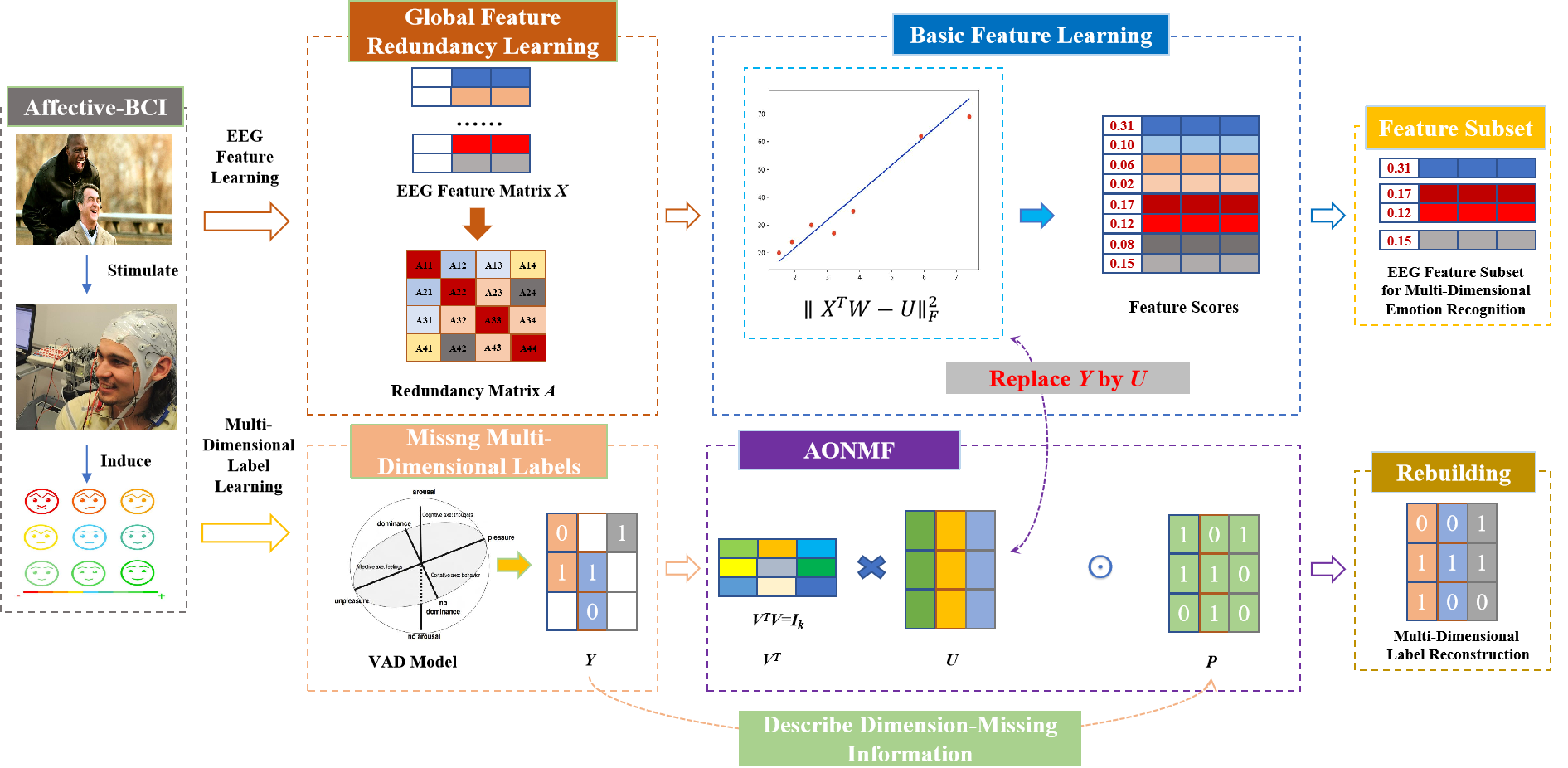}
\caption{REFS concludes the following three sections: (a) basic feature learning  with graph-based manifold regularization; (b) adaptive orthogonal non-negative matrix factorization (AONMF); and (c) global feature redundancy learning.}
\label{Framework_efsmmder}
\end{figure*}

\section{Related Works}\label{RW}
\subsection{EEG feature selection methods}
EEG feature selection techniques could be broadly categorized into three groups based on their interaction with learning models: filter, wrapper, and embedded methods \cite{zhang2019review}. Filter methods evaluate the importance of each feature according to specific criteria. The most informative EEG features, which score highly, are then selected. Examples of such techniques include information gain \cite{chen2015electroencephalogram}, mRMR \cite{wang2011eeg,lin2024decoding}, ReliefF \cite{zhang2016relieff}, etc. However, the filter methods may overlook potentially important features that, when combined with other features, could improve discrimination. This is because the learning models are not considered during the feature selection process \cite{saeys2007review}. Wrapper methods, by contrast, "wrap" feature subsets into candidate EEG feature sets using random or sequential search strategies. A learning model then evaluates the performance of these candidate subsets. Several widely used wrapper methods, such as the evolutionary computation algorithm \cite{nakisa2018evolutionary} and the ReliefF-based genetic algorithm \cite{kong2017rega}, have been proposed or applied to the EEG-based affective computing task. However, the computational cost of wrapper methods is often much higher than filter methods due to the iterative strategy of candidate feature subset search \cite{bolon2019ensembles}.

To overcome the limitations of the above feature selection approaches, embedded methods have been introduced \cite{hao2025embedded}. These methods integrate feature selection into the model learning process, evaluating the relative importance of each EEG feature while simultaneously optimizing learning models. The most frequently used statistical model for embedded feature selection is least squares regression (LSR), which is recognized for its statistical completeness and efficiency in data analysis \cite{nie2010RFS, chen2018semi, xu2021general, wu2019FSOR}. LSR-based feature selection algorithms aim to learn a projection matrix $W$ and rank feature significance using $\{\|{w}^{1}\|_{2},..., \|{w}^{d}\|_{2}|\}$ \cite{yang2019unsupervised}. Discriminative EEG feature subsets can then be formed based on their significance. Several popular LSR-based methods, including robust feature selection(RFS)\cite{nie2010RFS}, feature selection with orthogonal regression (FSOR) \cite{xu2020fsorer}, global redundancy minimization in orthogonal regression (GRMOR)\cite{GRMOR2021taffc}, weighted self-expression learning (WSEL)\cite{xu2024wsel}, have been developed or proposed for constructing EEG emotional feature subsets.

\section{Problem formulation}\label{Pf}
The proposed REFS framework is given as follows:
\begin{equation}
\min _{W, U, V} F(X, W, U)+\lambda E(Y, U, V)+\gamma \Omega(W)
\end{equation}
where $W \in \mathbb{R}^{d \times k}$ and $U \in \mathbb{R}^{n \times k}$ denote the projection matrix and latent structure matrix, respectively. $X \in \mathbb{R}^{d \times n}$ represents EEG emotional feature data, while $Y \in\{0,1\}^{n\times k}$ is the multi-dimensional emotional label matrix. $d$, $n$, and $k$ correspond to the number of features, instances, and dimensions, respectively. $\lambda$ and $\gamma$ denote tradeoff parameters. The basic feature learning function with graph-based manifold regularization, adaptive orthogonal non-negative matrix factorization, and global feature redundancy learning function are represented by $F$, $E$, and $\Omega$, respectively. The subsequent subsections will define $F$, $E$, and $\Omega$.

\subsection{Basic feature learning with graph-based manifold regularization}
Least square regression is utilized to model of the relationship between the EEG feature data $X$ and the latent structure matrix $U$. The basic feature learning function can be formulated as follows:
\begin{equation}
\begin{aligned}
\label{equ_framework_lsr} \left\|X^T W - U\right\|_F^2 + \delta\|W\|_{2,1} \text { s.t. } U \geq 0
\end{aligned}
\end{equation}
where the projection matrix $W$ is regularized by the $l_{2,1}$-norm to facilitate EEG emotional feature selection, and $\delta$ ($\delta > 0$) is a tradeoff parameter. 

The Frobenius norm and $l_{2,1}$ norm of a matrix $X$ are represented as $\|X\|_{F}=\sqrt{\sum_{i=1}^{d} \sum_{j=1}^{n} x_{i, j}^{2}}=\sqrt{t r\left(X^{T} X\right)}$ and $\|X\|_{2,1}=\sum_{i=1}^{d} \sqrt{\sum_{j=1}^{n} x_{i, j}^2}=\sum_{i=1}^{d}\left\|x_{i,:}\right\|_2$.

Then, a graph-based manifold regularizer is adopted to retain the consistency of local geometric structures between the original EEG feature space and the latent structure space \cite{jian2016multi}. According to spectral graph theory, two data points in the EEG feature matrix $X$ that exhibit high correlation imply a high similarity between their latent representations in the latent structure matrix $U$, which can be formulated as:
\begin{equation}
\label{equ_gbmr}
\begin{aligned}
\frac{1}{2} \sum_{i=1}^n \sum_{j=1}^n S_{i j}\left\|U_{i.}-U_{j.} \right\|_2^2 &=\operatorname{Tr}\left(U^T(G-S) U\right)\\
&=\operatorname{Tr}\left(U^T L_{X} U\right)
\end{aligned}
\end{equation}
where $S$, $G$, and $L_X \in \mathbb{R}^{n \times n}$ denote the affinity graph for the original EEG feature matrix $X$, a diagonal matrix with $G_{ii} = \sum_{j=1}^{n} S_{ij}$, and the graph Laplacian matrix for $X$, respectively. $L_X$ is computed as $L_X = G - S$. To acquire the affinity graph matrix, we utilize a heat kernel function. The element $S_{i j}$ of $S$ is defined as follows:
\begin{equation}
S_{i j}=\left\{\begin{array}{lc}
\exp \left(-\frac{\left\|\bm{x}_{.i}-\bm{x}_{.j}\right\|^{2}}{\sigma^{2}}\right) &\bm{x}_{.i} \in \mathcal{N}_{q}\left(\bm{x}_{.j}\right) \\ &\text { or } \bm{x}_{.j} \in \mathcal{N}_{q}\left(\bm{x}_{.i}\right) \\
0 & \text { otherwise }
\end{array}\right.
\end{equation}
where $\mathcal{N}_q\left(\bm{x}_{.j}\right)$ represents the set of the top $q$ nearest neighbors of the sample $\bm{x}_{.j}$, and $\sigma$ denotes the parameter of the heat kernel mode.

Finally, by combining \eqref{equ_framework_lsr} and \eqref{equ_gbmr}, the formulation of the term $F$ can be expressed as follows:
\begin{equation}
\begin{aligned}
\label{equ_framework_R}
F(X, W, U)= &\left\|X^T W - U\right\|_F^2+ \eta \operatorname{Tr}\left( U^T L_X U\right) +\delta\|W\|_{2,1} \\
&\text { s.t. } U \geq 0
\end{aligned}
\end{equation}
where $\eta$ and $\delta$ denote tradeoff parameters.

\subsection{Adaptive orthogonal non-negative matrix factorization}
The traditional non-negative matrix factorization model decomposes a matrix into two non-negative matrices. To preserve more local structural information and reduce the impact of outliers, an orthogonal constraint $V^{T} V=I_{k}$ is imposed on the coefficient matrix $V$. The cost function for non-negative matrix factorization with orthogonal constraints is formulated as follows:
\begin{equation}
\begin{aligned}
\label{equ_framework_se}
&\min _{U,V} \|Y- U V^T\|_{F}^{2}  \\
&\text { s.t. } U \geq 0, V\geq 0,V^{T} V=I_{k}
\end{aligned}
\end{equation}
where $U$ and $V \in \mathbb{R}^{k \times k}$ denote the latent structure matrix and a coefficient matrix of $Y$, respectively.

To represent the dimension-missing information, the element $P_{i,j}$ of a weighting mask matrix $P$ is defined as:
\begin{equation}
P_{i,j}= \begin{cases}1 & \text { if } i \text {-th instance exists in } j \text {-th dimensional label; } \\ 0 & \text { otherwise. }\end{cases}
\end{equation}

Then, $P$ is incorporated into the orthogonal non-negative matrix factorization function as follows:
\begin{equation}
\begin{aligned}
\label{equ_framework_se}
&E(Y, U, V) =  \|P \odot (Y- U V^T)\|_{F}^{2}\\
&\text { s.t. } U \geq 0, V\geq 0,V^{T} V=I_{k}
\end{aligned}
\end{equation}
where $\odot$ denotes the Hadamard product. By leveraging second-order and high-order label correlations among the multi-dimensional emotional labels, each dimension with missing values can be rebuilt via all other dimensions.

\subsection{Global feature redundancy learning}
Additionally, a global feature redundancy matrix $A$ is introduced to analyze the redundancy among all features. The elements of $A$ are defined as follows:
\begin{equation}
\label{equ_red_matrix}
A_{i, j}=\left(O_{i, j}\right)^{2}=\left(\frac{\bm{f}_{i}^{T} \bm{f}_{j}}{\left\|\bm{f}_{i}\right\|\left\|\bm{f}_{j}\right\|}\right)^{2}
\end{equation}
where $\bm{f}_{i}\in {\mathbb{R}^{n\times 1}}$ and $\bm{f}_{j}\in {\mathbb{R}^{n\times 1}}$ represent the $i$-th and $j$-th centralized representation of two kinds of EEG features $\bm{x}_{i}$ and $\bm{x}_{j}$ ($i,j=1,2,...,\textit{d}$). $\bm{f}_{i}$ and $\bm{f}_{j}$ are defined as follows:
\begin{equation}\left\{\begin{array}{l}
\bm{f}_{i}=H \bm{x}_{i}^{T} \\
\bm{f}_{j}=H \bm{x}_{j}^{T}
\end{array}\right.\end{equation}
where $H={{I}_{\text{n}}}-\frac{1}{n}{{\bm{1}}_{n}}{{\bm{1}}_{n}}^{T}$. \eqref{equ_red_matrix} can be changed to
\begin{equation}
O=Z F^{T} F Z=(F Z)^{T} F Z
\end{equation}
where $F=[\bm{f}_1, \bm{f}_2, ..., \bm{f}_d]$. Define $Z$ as a diagonal matrix where each diagonal element is given by $Z_{i, i}=\frac{1}{\left\|\bm{f}_{i}\right\|}$ ($i=1,2,...,\textit{d}$). 

Since the matrix $O$ is positive semi-definite and $A = O \odot O$, $A$ is non-negative and positive semi-definite matrix \cite{wang2015feature}. Consequently, the global feature redundancy learning function is formulated as follows:
\begin{equation}
\begin{aligned}
\label{equ_framework_theta}
\Omega(W) = \operatorname{Tr}\left(W^T A W\right)
\end{aligned}
\end{equation}

\subsection{The final objective function of REFS}
By combining \eqref{equ_framework_R}, \eqref{equ_framework_se}, and \eqref{equ_framework_theta}, the proposed REFS model is summarized as follows:
\begin{equation}
\begin{aligned}
\label{equ_framework}
\min _{W, V, U}&\left\|X^T W - U\right\|_F^2+\lambda\|P \odot(Y- U V^T)\|_F^2 \\
&+\eta \operatorname{Tr}\left( U^T L_X U\right)+\mu \operatorname{Tr}\left(W^T A W\right)+\delta\|W\|_{2,1} \\
&\text { s.t. } U \geq 0, V \geq 0,V^{T} V=I_{k}
\end{aligned}
\end{equation}
where $\lambda$, $\eta$, $\mu$, and $\delta$ are regularization parameters. The flowchart of REFS is illustrated in Fig.~\ref{Framework_efsmmder}.

\section{Optimization Strategy} \label{Optimization Strategy}
The alternatively iterative optimization strategy is employed to derive solutions for the three variables ($W$, $V$, and $U$) in \eqref{equ_framework}. The strategy is described as follows:

\subsection{Update $W$ by fixing $U$, $V$}
When $U$ and $V$ are fixed and irrelevant terms are omitted, the following function for $W$ is obtained:
\begin{equation}
\label{equ_w_l}
\mathcal{L}\left(W\right) = \left\|X^T W - U\right\|_F^2+\mu \operatorname{Tr}\left(W^T A W\right)+\delta\|W\|_{2,1}
\end{equation}

By taking the partial derivative of $\mathcal{L}\left(W\right)$ with respect to $W$, we obtain:
\begin{equation}
\label{equ_w_pd}
\frac{\partial \mathcal{L}\left(W\right)}{\partial W} = 2XX^TW-2XU+2\mu AW+2\delta DW
\end{equation}
where $D$ is a diagonal matrix, and its elements are computed as $D_{ii}=\frac{1}{2 \sqrt{W_i^T W_i+\epsilon}}$ (where $\epsilon \rightarrow 0$).

Set $\frac{\partial \mathcal{L}\left(W\right)}{\partial W} = 0$. Consequently, the optimal solution for $W$ can be updated as follows:
\begin{equation}
\label{equ_w_sol}
W = (X X^T+\mu A+\delta D)^{-1}(XU)
\end{equation}

\subsection{Update $U$ by fixing $W$, $V$}
When $W$ and $V$ are fixed, by introducing a Lagrange multiplier $\mathbf{\Psi}$ for $U \geq 0$, the Lagrange function is given by:
\begin{equation}
\begin{aligned}
\label{equ_u_pd}
\mathcal{L}\left(U\right)=&\left\|X^T W - U\right\|_F^2+\lambda\|P \odot(Y- U V^T)\|_F^2\\
&+\eta \operatorname{Tr}\left(U^T L_X U\right) +\operatorname{Tr}\left(\mathbf{\Psi}^T U\right) \\
\end{aligned}
\end{equation}

Then, the partial derivative of $\mathcal{L}(U)$ with respect to $U$ in \eqref{equ_u_pd} is computed as follows:
\begin{equation}
\begin{aligned}
\frac{\partial \mathcal{L}\left(U\right)}{\partial U}=&-2 X^T W + 2 U + 2 \lambda (P\odot(U V^T-Y)) V \\
&+2\eta L_X U+\mathbf{\Psi}
\end{aligned}
\end{equation}

Via the Karush-Kuhn-Tucker (KKT) complementary condition $\mathbf{\Psi}_{ij}U_{ij} = 0$, the update rule for $U$ is:
\begin{equation}
\label{equ_u_sol}
U \leftarrow U \odot \frac{X^T W+\lambda (P \odot Y)V}{U +\lambda (P\odot(U V^T))V+\eta L_X U}
\end{equation}

\subsection{Update $V$ by fixing $U$, $W$}
When $W$ and $U$ are fixed, by introducing Lagrange multipliers $\mathbf{\Phi}$ for $V \geq 0$ and $\xi$ for $V^{T} V=I_{k}$, the Lagrange function is given by:
\begin{equation}
\begin{aligned}
\label{equ_v_pd}
\mathcal{L}\left(V\right)=&\lambda\|P \odot(Y- U V^T)\|_F^2 + \xi \|V^{T} V- I\|_F^2  \\
&+\operatorname{Tr}\left(\mathbf{\Phi}^T V\right)
\end{aligned}
\end{equation}

Then, the partial derivative of $\mathcal{L}\left(V\right)$ with respect to $V$ in \eqref{equ_v_pd} is calculated as:
\begin{equation}
\begin{aligned}
\frac{\partial \mathcal{L}\left(V\right)}{\partial V}=& 2 \lambda (P^T\odot(V U^T- Y^T)) U \\
& + \xi(4 V V^T V- 4V) + \mathbf{\Phi}\\
\end{aligned}
\end{equation}

Via the KKT complementary condition $\mathbf{\Phi}_{ij}V_{ij} = 0$, the update rule for $V$ is:
\begin{equation}
\label{equ_v_sol}
V \leftarrow V \odot \frac{\lambda (P^T \odot Y^T)U + 2\xi V}{\lambda (P^T\odot(V U^T))U+ 2 \xi V V^T V}
\end{equation}

\begin{algorithm}[h]
\caption{Robust EEG feature selection with missing multi-dimensional annotation for emotion recognition}
\label{REFS}
\begin{algorithmic}[1]
\Require $X\in {\mathbb{R}^{\text{d}\times n}}$: EEG data matrix. $Y\in {\mathbb{R}^{n\times k}}$: missing multi-dimensional emotional label matrix. $P\in {\mathbb{R}^{n\times k}}$: indicator matrix for dimension-missing information.
\Ensure Return ranked EEG features.
\State Initial $W$ , $U$, and $V$ randomly.
\Repeat
\State Update $D$ via $D_{i i}=\frac{1}{2 \sqrt{W_i^T W_i+\epsilon}}$;
\State Update $W$ via $W = (X X^T+\mu A+\delta D)^{-1}(XU)$;
\State Update $U$ via \eqref{equ_u_sol};
\State Update $V$ via \eqref{equ_v_sol};
\Until{Convergence;}
\State \Return $W$ for EEG feature selection.
\State Rank the EEG features via $ \|\bm{w}_{i}\|_{2}$;
\end{algorithmic}
\end{algorithm}

Algorithm~\ref{REFS} outlines the detailed optimization steps for the objective \eqref{equ_framework}. The scores for EEG features are computing using $ \|\bm{w}_{i}\|_{2}$ and the informative and non-redundant EEG features with highest scores are chosen to construct optimal feature subsets for emotion recognition.

\section{Experiments and Discussion} \label{Experimental Details}

\subsection{Datasets}
Simulation experiments were conducted on three publicly available EEG datasets (HDED \cite{GRMOR2021taffc}, DEAP \cite{koelstra2011deap}, and DREAMER \cite{DREAMER2018jbhi}) to evaluate the reliability of REFS. all datasets utilize the VAD paradigm to record the emotional states of subjects. We employed a band-pass filter with a cutoff frequency between 1 to 50 Hz to minimize noise in the original EEG signals. Following this, independent component analysis was implemented to mitigate artifacts from muscular and eye movements. Each entire trial was regarded as a single sample for EEG feature extraction. As stated differently, trials were not subdivided into smaller segments to increase the sample size for the experiments.

\subsection{EEG feature extraction}
According to previous EEG feature extraction research for emotion recognition \cite{jenke2014taffc,xu2020fsorer}, thirteen kinds of EEG features was extracted to construct EEG feature matrix, which includes functional connectivity, differential entropy, C0 complexity, higher-order crossing , non-stationary index, Shannon entropy, rational asymmetry, differential asymmetry, absolute power, spectral entropy,  the ratio of absolute power in the $\beta$ band to that in the $\theta$ band, the amplitude of the Hilbert transform of intrinsic mode functions, and the instantaneous phase of the Hilbert transform of intrinsic mode functions. Detailed descriptions of the thirteen feature types are available in \cite{Duan2013DE,jenke2014taffc,GRMOR2021taffc}. The total dimensions of the thirteen feature types are 651 for DREAMER, 1756 for DEAP, and 3150 for HDED.

\subsection{Experimental details}
Thirteen feature selection methods were compared to thoroughly evaluate the validity and reliability of REFS. These methods include:

(1) Two popular embedded EEG emotional feature selection methods: FSOR \cite{xu2020fsorer} and GRMOR \cite{GRMOR2021taffc}.

(2) Seven popular multi-label feature selection approaches: PMU \cite{lee2013feature}, FIMF \cite{lee2015fast}, SCLS \cite{lee2017scls}, MDFS \cite{zhang2019manifold}, GRRO \cite{zhang2020multilabel}, MGFS \cite{hashemi2020mgfs}, and MFS-MCDM \cite{hashemi2020mfs}. 

(3) Four incomplete multi-label feature selection approaches: MLMLFS \cite{zhu2018multi}, FSMML \cite{yin2024feature}, WSEL \cite{xu2024wsel}, and WFDP \cite{dai2025multi}.

For each emotional dimension, the VAD paradigm based self-assessed scores were categorized into low and high classes via a threshold value of five. Multi-label k-nearest neighbor (ML-KNN) \cite{ZHANG2007MLKNN} was adopted as the base classifier for multi-dimensional emotion recognition, with the number of nearest neighbors and smoothing parameter set to 10 and 1, respectively. Additionally, six performance metrics were implemented to compare the performance. These performance metrics comprise two label-based evaluation indices (macro-F1 (MA), micro-F1 (MI)) and four instance-based evaluation indices (ranking loss (RL), average precision (AP), coverage (CV), and hamming loss (HL)). Detailed information on these metrics can be found in \cite{zhang2019manifold}. A cross-subject experimental setup was employed, where 70\% of the subjects were randomly selected to form the training set, and the remaining 30\% of the subjects served as the test set. To minimize bias, 50 separate and independent experiments were conducted, and the average result was considered as the final measure of emotion recognition.

%\begin{table}[!t]
%\caption{The accurate recovery ratio(\%) of missing labels.}\label{tab:acmkd}
%\begin{center}
%{
%\begin{tabular}{ccc}
%\hline\hline
%\multicolumn{1}{l}{\multirow{2}*{Missing ratio}}         & \multicolumn{2}{c}{\multirow{1}*{Accurate recovery ratio}} \\  \cline{2-3}
%                         & DEAP        & DREAMER               \\\hline
%10\%                     & 69.09          & 69.58              \\
%20\%                     & 67.58          & 68.76              \\
%30\%                     & 65.49          & 66.86              \\
%40\%                     & 65.12          & 65.82              \\
%50\%                     & 63.23          & 64.52              \\ \hline\hline
%\end{tabular}}
%\end{center}
%\end{table}

\begin{figure}[!t]
\centering
\includegraphics[width=0.495\textwidth]{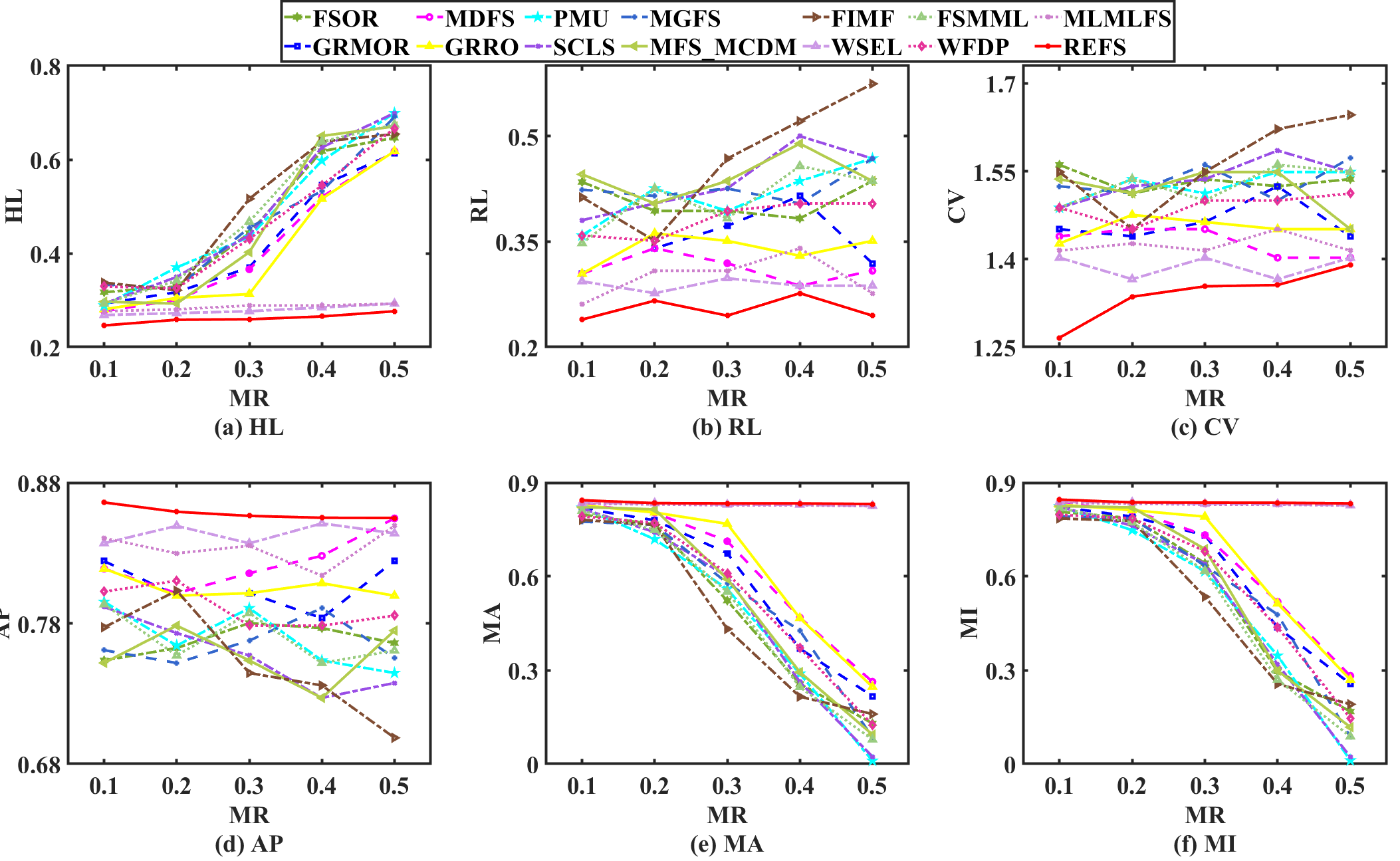}
\caption{Multi-dimensional emotion recognition performance of various feature ratios (FR) on DREAMER.}\label{Results_index_dreamer}
\end{figure}

\begin{figure}[!t]
\centering
\includegraphics[width=0.495\textwidth]{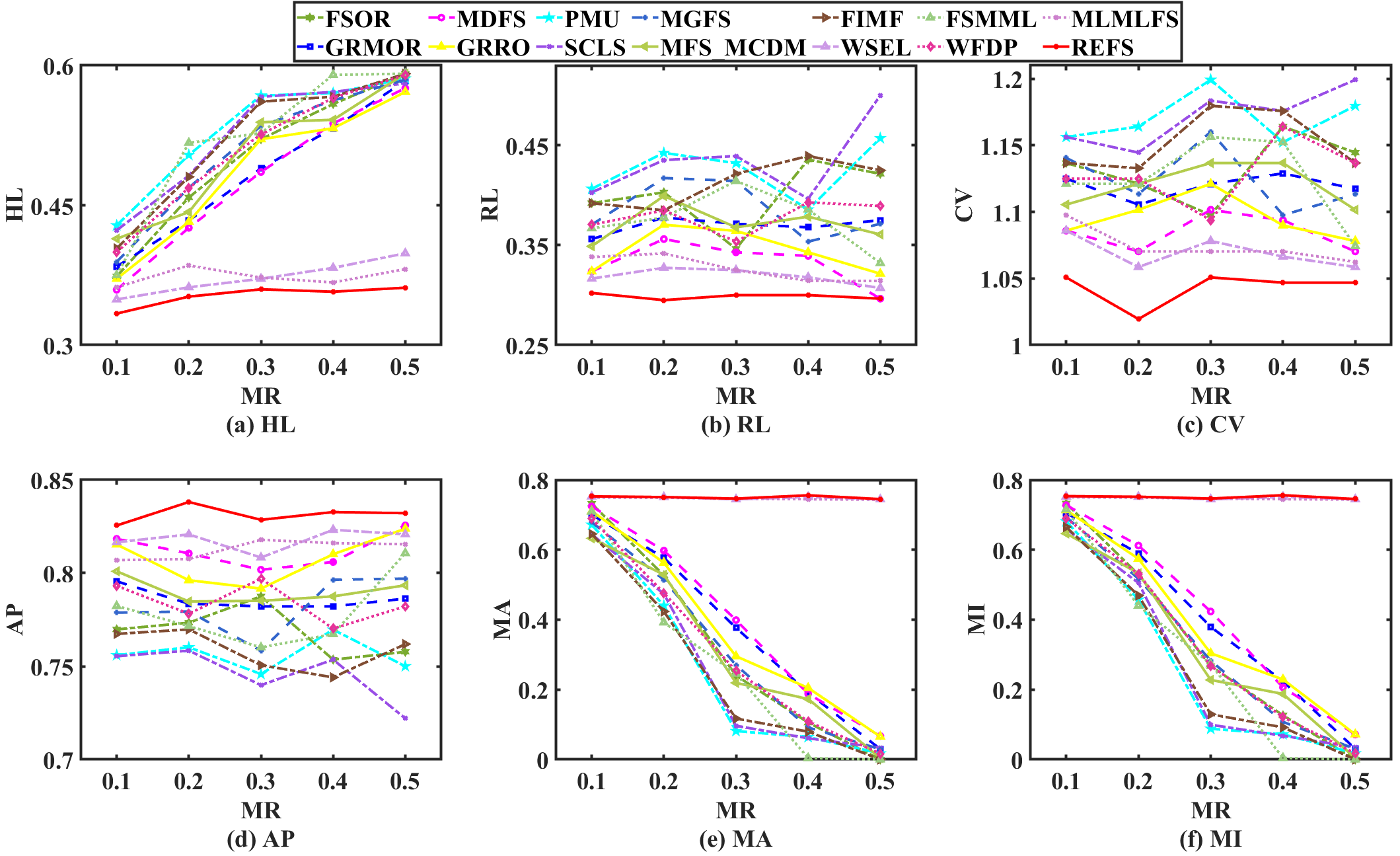}
\caption{Multi-dimensional emotion recognition performance of various feature ratios on DEAP.}\label{Results_index_deap}
\end{figure}

\begin{figure}[!t]
\centering
\includegraphics[width=0.495\textwidth]{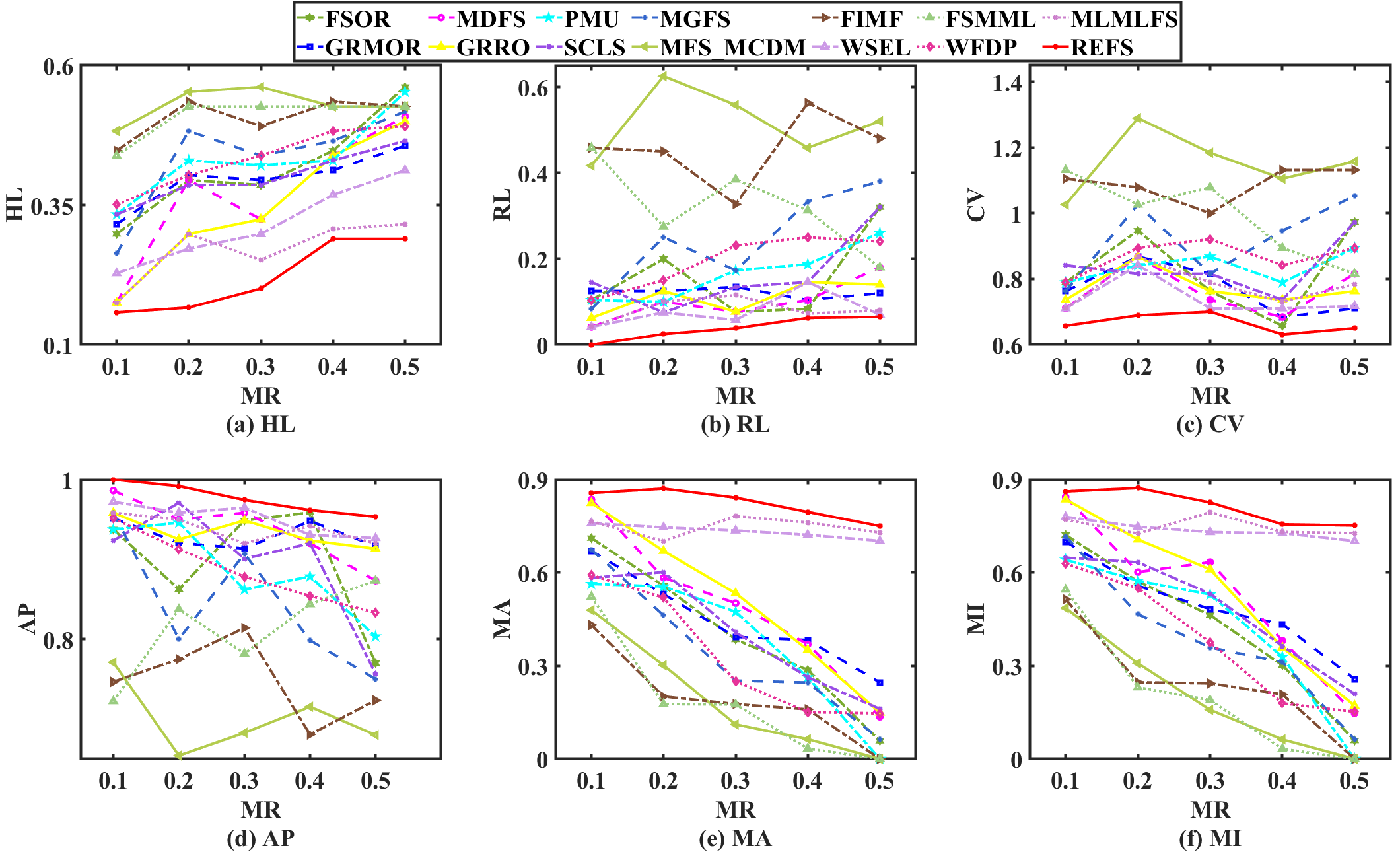}
\caption{Multi-dimensional emotion recognition performance of various feature ratios on HDED.}\label{Results_index_hded}
\end{figure}

\subsection{Comparison with baseline methods}
We adopted the strategy outlined in \cite{liu2018late} to simulate missing multi-label scenarios by removing varying percentages of emotional labels from each dimension to generate missing data. The missing ratio ranged from 10\% to 50\%, in increments of 10\%. Using the feature selection approaches, approximately 10\% of original EEG features were chosen. The tradeoff parameters of REFS were tuned accordingly from $10^{- 3}$ to $10^{3}$ with a step of $10^{1}$.

Fig. 3-5 present the comparative results of ten methods applied to the three datasets. The vertical axis shows the results of each evaluation indices, and the horizontal axis denotes the missing ratios of multi-dimensional emotional labels. As illustrated in Fig. 3(a-c), Fig. 4(a-c), and Fig. 5(a-c), performance of multi-dimensional emotion recognition improves as the values of evaluation metrics decrease. Conversely, as shown in Fig. 3(d-f), Fig. 4(d-f), and Fig. 5(d-f), performance improves as the values increase. Overall, the results presented in Fig. 3-5 demonstrate that REFS has superior performance across various missing ratios.

The Friedman test was utilized to further compare the multi-dimensional emotion recognition results of the ten feature selection methods, with the significance level set at $\alpha=0.05$. The results of the statistical significance test are shown in Table~\ref{tab:friedman}. It can be observed that the null hypothesis is rejected, indicating that there are significant differences in the performance of the fourteen feature selection methods for missing multi-dimensional emotion recognition. 

\begin{table}[!t]\small
\begin{center}
{
\begin{tabular}{lcc}
\hline\hline
Evaluation metric     & $F_{F}$             & Critical value  \\ \hline
Ranking loss          & 14.921              & \multicolumn{1}{c}{\multirow{6}*{$\approx$ 2.064}}             \\
Coverage              & 13.133             &            \\
Hamming loss          & 16.067             &            \\
Average precision     & 12.945             &           \\
Macro-F1              & 19.328            &            \\
Micro-F1              & 18.073             &            \\
\hline\hline
\end{tabular}}
\caption{The Friedman test results ($\alpha = 0.05$).}\label{tab:friedman}
\end{center}%\vskip -5pt
\end{table}

%\begin{figure}[!t]
%\centering
%\subfigure[Hamming loss]{\label{HL_neme}\includegraphics[width=0.23\textwidth]{HL_neme}}
%\hspace{0.0cm}
%\subfigure[Coverage]{\label{CV_neme}\includegraphics[width=0.23\textwidth]{CV_neme}}
%\hspace{0.0cm}
%\subfigure[Ranking loss]{\label{RL_neme}\includegraphics[width=0.23\textwidth]{RL_neme}}
%\hspace{0.0cm}
%\subfigure[Average precision]{\label{AP_neme}\includegraphics[width=0.23\textwidth]{AP_neme}}
%\hspace{0.0cm}
%\subfigure[Macro-F1]{\label{MA_neme}\includegraphics[width=0.23\textwidth]{MA_neme}}
%\hspace{0.0cm}
%\subfigure[Micro-F1]{\label{MI_neme}\includegraphics[width=0.23\textwidth]{MI_neme}}
%\caption{The Nemenyi test results ($\alpha =0.05$).} \label{Results_neme}
%\end{figure}

\subsection{Recovery results of missing labels}
%\begin{figure}[!t]
%\centering
%\includegraphics[width=0.3\textwidth]{ACR}
%\caption{The accurate recovery ratio(\%) results of missing labels.}
%\label{ACR}
%\end{figure}

\begin{figure}[!t]
\centering
\includegraphics[width=0.4\textwidth]{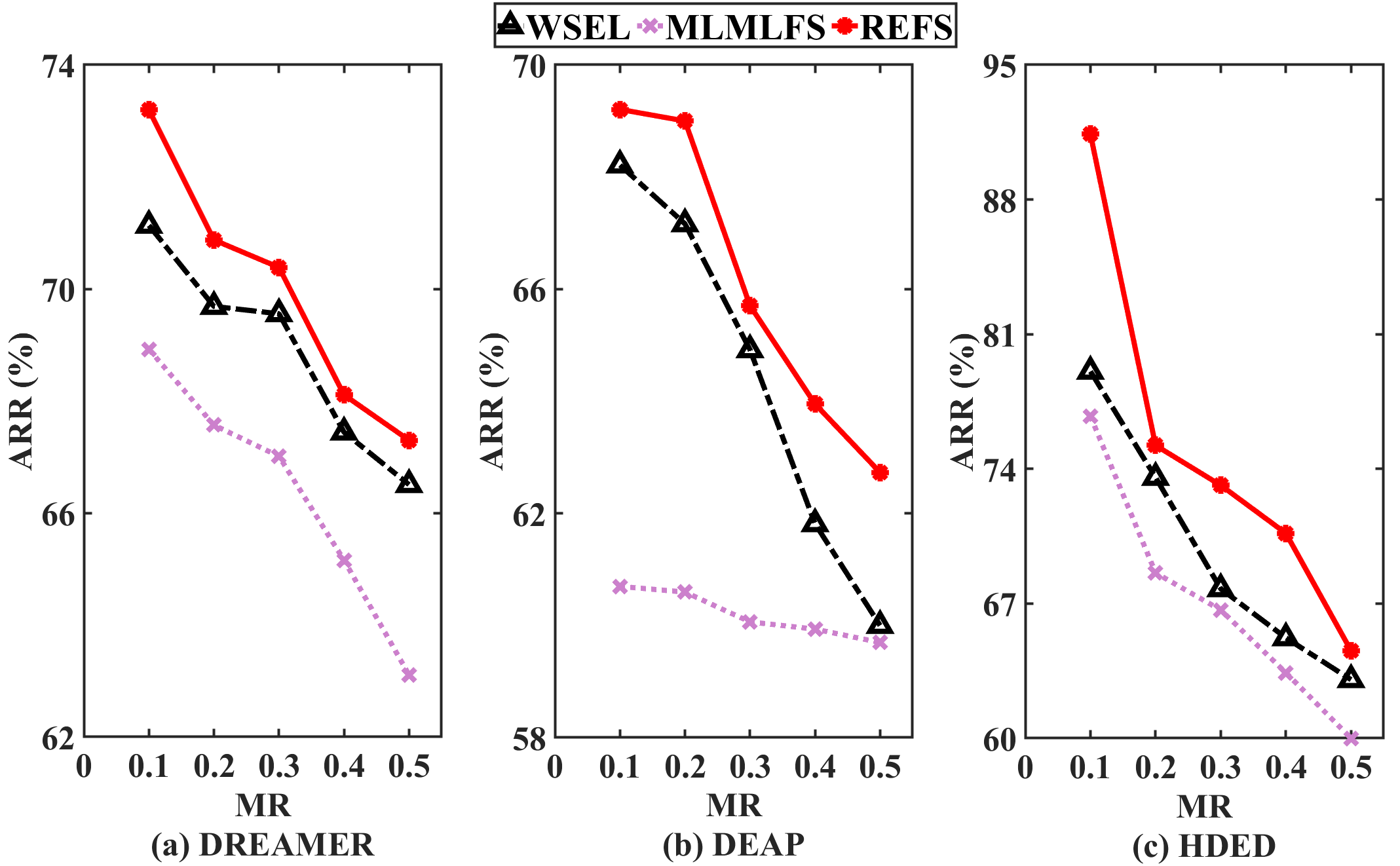}
\caption{Accurate recovery ratios (ARR) of missing labels.}\label{ACR}
\end{figure}

To further validate the effectiveness of the adaptive orthogonal non-negative matrix factorization module for label reconstruction, Fig.~\ref{ACR} summarizes the proportion of successfully recovered emotional labels at various missing rates. As shown in Fig.~\ref{ACR}, the proposed method effectively recovers over 60\% of the missing labels using this module, thereby providing more comprehensive label information for the basic feature learning module.

\subsection{Ablation experiments}
Additionally, we conducted ablation experiments to assess the contributions of each module in REFS. Table~\ref{tab:acml} reveals that the weighted matrix and orthogonal constraint modules are crucial for recovering missing labels, as it effectively leverages high-order correlations within missing emotional labels. The other modules contribute to suppressing redundant information and maintaining local geometry structures.

\begin{table}[!t]\small
\begin{center}
{
\begin{tabular}{lcc||cc||cc}
\hline\hline
\multicolumn{1}{l}{\multirow{2}*{Methods}}  & \multicolumn{2}{c}{\multirow{1}*{DREAMER}}        & \multicolumn{2}{c}{\multirow{1}*{DEAP}}                  &  \multicolumn{2}{c}{\multirow{1}*{HDED}}  \\  \cline{2-7}
                  &  HL $\downarrow$       & AP$\uparrow$           &  HL $\downarrow$       & AP$\uparrow$   & HL $\downarrow$      & AP $\uparrow$               \\\hline
w/o WM           &0.41                  &0.79   & 0.42                   &0.78   & 0.27                       & 0.94                                         \\      
w/o OC           &0.42                  &0.77   & 0.45                   &0.79  & 0.29                       & 0.94                                            \\
w/o GFRL         &0.31                  &0.81  & 0.39                   &0.82  & 0.24                       & 0.97                                            \\
w/o GMR          &0.33                  &0.80  & 0.38                   &0.81  & 0.25                       & 0.96                                            \\
\textbf{REFS}   &\textbf{0.27}         &\textbf{0.85}  & \textbf{0.36}          &\textbf{0.84}  & \textbf{0.22}                       & \textbf{0.98}                             \\ \hline\hline
\end{tabular}}
\caption{The ablation experimental results (w/o, WM, OC, GFRL, and GMR denote without, weighted matrix, orthogonal constraints, global feature redundancy learning, and graph-based manifold regularizer, respectively).}\label{tab:acml}
\end{center}
\end{table}

\begin{figure}[!t]
\centering
\subfigure[$\lambda$]{\label{bar3_lambda}\includegraphics[width=0.16\textwidth]{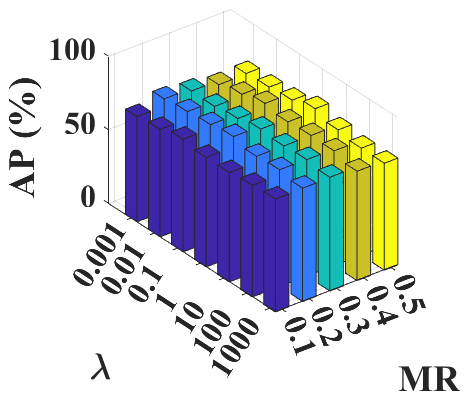}}
\hspace{0.0cm}
\subfigure[$\eta$]{\label{bar3_eta}\includegraphics[width=0.16\textwidth]{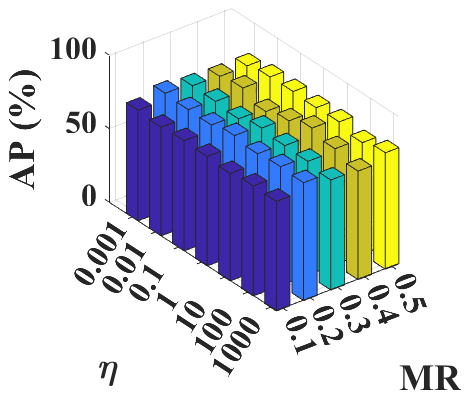}}
\hspace{0.0cm}
\subfigure[$\mu$]{\label{bar3_mu}\includegraphics[width=0.16\textwidth]{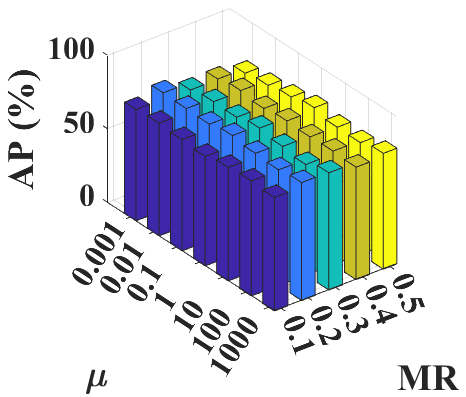}}
\hspace{0.0cm}
\subfigure[$\delta$]{\label{bar3_delta}\includegraphics[width=0.18\textwidth]{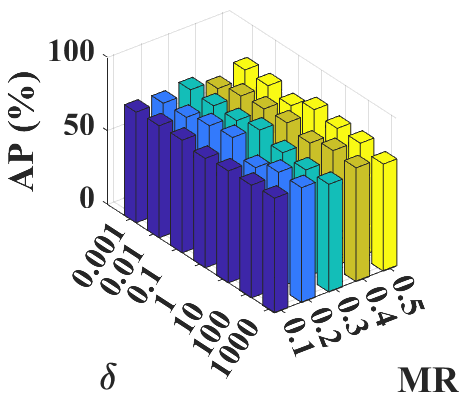}}
\caption{The parameter sensitivity results on DEAP.}
\label{Results_bar3_deap}
\end{figure}

\begin{table}[!t]\small
\begin{center}
{
\begin{tabular}{cccc}
\hline\hline
Methods           &HDED       &  DEAP       & DREAMER    \\\hline
MDFS              & 21.87      & 4.08       & 1.45         \\
GRRO              & 35.66      & 32.42      & 2.01       \\
PMU               & 102.24      & 134.19     & 7.54         \\
SCLS              & 28.49      & 27.51      & 1.73       \\
MGFS              & 1.27      & 0.45       & 0.34        \\
MFS\_MCDM         & 3.91      & 0.65       & 0.18          \\
FIMF              & 0.09      & 0.03       & 0.01         \\
FSOR              & 21.34      & 369.05     & 46.42        \\
GRMOR             & 33.51      & 146.22     & 27.47          \\
WSEL              & 17.60      & 3.98       & 0.99               \\
FSMML             & 6497.86      & 12731.25           & 4336.28               \\
WFDP              & 1.10      & 32.07      & 7.12               \\
MLMLFS            & 0.38      & 4.20           & 0.43                \\
\textbf{REFS}     & \textbf{17.31}      & \textbf{9.26}          & \textbf{0.91}        \\ \hline\hline
\end{tabular}}
\caption{The comparison of computational time (seconds).}\label{tab:avetime}
\end{center}
\end{table}

\subsection{Parameter sensitivity analysis}
This section presents a parameter sensitivity analysis of REFS. In each analysis, one parameter was fixed at 0.1 while the others were varied. Due to space constraints, only the parameter sensitivity results for DEAP are presented. The 3D histogram in Fig.~\ref{Results_bar3_deap} quantifies AP stability. As illustrated in Fig.~\ref{Results_bar3_deap}, AP remains relatively stable as parameters vary, indicating that REFS is not highly sensitive to the parameters.

\subsection{Computational cost analysis}
The computational time of all methods was compared. The implementation was carried out in MATLAB (MathWorks Inc., Novi, MI, USA) and executed on a computer running Microsoft Windows $11 \times 64$, equipped with an Intel Core i5-12400HQ 2.5 GHz CPU and 16 GB of RAM. The average computational time results are presented in Table~\ref{tab:avetime}. As shown in Table~\ref{tab:avetime} and Fig. 3-5, REFS delivers optimal performance with a relatively low computational cost.

\begin{figure}[!t]
\centering
\subfigure[DREAMER]{\label{Conv_dreamer}\includegraphics[width=0.13\textwidth]{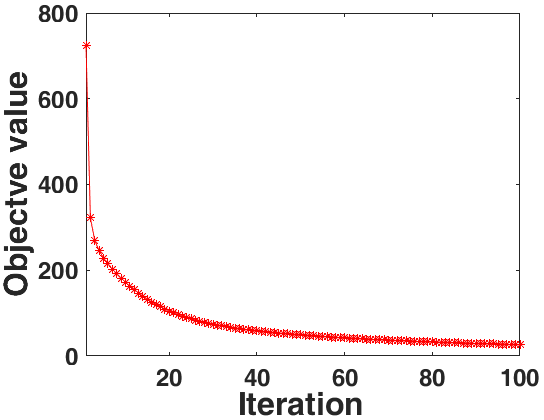}}
\hspace{0.0cm}
\subfigure[DEAP]{\label{Conv_deap}\includegraphics[width=0.13\textwidth]{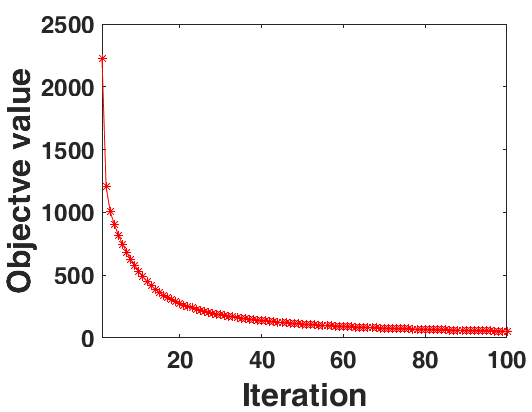}}
\hspace{0.0cm}
\subfigure[HDED]{\label{Conv_hded}\includegraphics[width=0.13\textwidth]{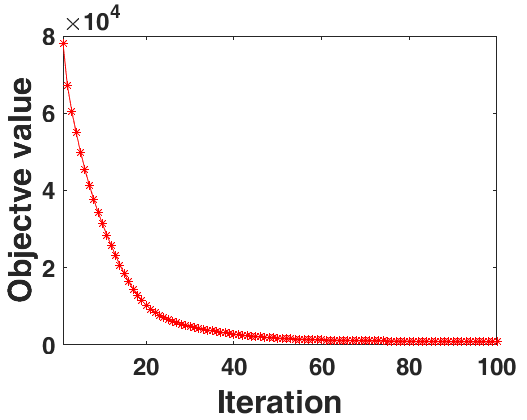}}
\caption{Convergence verification of REFS.}
\label{conv_res}
\end{figure}

\subsection{Convergence analysis}
Furthermore, a convergence speed analysis of the proposed iterative update algorithm was conducted. Fig.~\ref{conv_res} displays the convergence curves of the objective values. The regularization parameters ($\lambda$, $\eta$, $\mu$, and $\delta$) were uniformly set to 10. As demonstrated in Fig.~\ref{conv_res}, the REFS algorithm converges quickly within a few iterations, highlighting the effectiveness of our iterative optimization strategy.

\section{Conclusions}\label{Conclusion}
This paper presents a novel EEG feature selection model designed to handle partially missing labels for recognizing multi-dimensional emotions. REFS could select informative EEG features and recover missing labels by integrating adaptive orthogonal non-negative matrix factorization and global feature redundancy learning within a least squares regression framework. Additionally, a straightforward yet effective iterative optimization algorithm is introduced to solve the optimization problem of REFS. Experimental results indicate that REFS significantly outperforms thirteen advanced methods.

%Although the proposed method is able to recover missing emotion labels to some extent (with a recovery rate of over 60\%), the overall performance remains unsatisfactory, particularly in real-world affective computing applications based on brain-computer interface systems. Improving the recovery rate of missing emotion labels is an important challenge. In the future, we will focus on developing new strategies aimed at enhancing the recovery rate of missing emotion labels in the context of multi-dimensional emotion recognition.

\bibliography{aaai2026}

\end{document}